\documentclass[journal=ancac3,manuscript=article]{achemso}
\usepackage[version=3]{mhchem} 
\usepackage{graphicx}  
\usepackage{dcolumn}   
\usepackage{bm}        
\usepackage{amssymb}   
\usepackage{xcolor}
\usepackage[colorinlistoftodos]{todonotes} 

\author{Anouar Benali}
\affiliation{Argonne Leadership Computing Facility, Argonne National Laboratory, Argonne, Illinois 60439 USA}

\author{Luke Shulenburger}
\affiliation{HEDP Theory Department, Sandia National Laboratories, Albuquerque, New Mexico 87185 USA}

\author{Jaron T. Krogel}
\affiliation{Materials Science and Technology Division, Oak Ridge National Laboratory, Oak Ridge, Tenessee 37831 USA}

\author{Xiaoliang Zhong}
\affiliation{Material Science Division, Argonne National Laboratory, Argonne, Illinois 60439 USA}

\author{Paul R. C. Kent}
\affiliation{Center for Nanophase Materials Sciences and Computer Science and Mathematics Division, Oak Ridge National Laboratory, Oak Ridge, Tenessee 37831 USA}

\author{Olle Heinonen}
\email{heinonen@anl.gov}
\affiliation{Material Science Division, Argonne National Laboratory, Argonne, Illinois 60439 USA}
\alsoaffiliation{Northwestern-Argonne Institute for Science and Engineering, Northwestern University, 2145 Sheridan Rd., Evanston, Illinois 60208, USA}

\title{Quantum Monte Carlo analysis of a charge ordered insulating antiferromagnet: the Ti$_4$O$_7$ Magn\'eli phase}
\begin{document}

\begin{abstract}
The Magn\'eli phase Ti$_4$O$_7$ is an important transition metal oxide with a wide range of applications because of its interplay between charge, spin, and lattice degrees of freedom. At low temperatures, it has non-trivial magnetic states very close in energy, driven by electronic exchange and correlation interactions. We have examined three low-lying states, one ferromagnetic and two antiferromagnetic, and calculated their energies as well as Ti spin moment distributions using highly accurate Quantum Monte Carlo methods. We compare our results to those obtained from density functional theory-based methods that include approximate corrections for exchange and correlation. Our results confirm the nature of the states and their ordering in energy, 
as compared with density-functional theory methods. However, the energy differences and spin distributions differ. A detailed analysis suggests that non-local exchange-correlation functionals, in addition to other approximations such as LDA+U to account for correlations, are needed to simultaneously obtain better estimates for spin moments, distributions, energy differences and energy gaps.
\end{abstract}

\section{Introduction}
Electronic correlations play an important role in compounds containing elements with partially filled d- and f-shells, such as transition metals in rows three and four of the periodic table. Correlation effects are responsible for a number of phenomena of intense interest during the past several decades\cite{ZaanenPRL1985,Anisimov1997,AnisimovJOP1997,DagottoScience2005}. In particular, the oxides of 3d transition metals, such as Mn, Ti, Fe, Co, Ni, and Zn, exhibit strong electronic correlations because of their rather well localized d-orbitals with many different oxidation states. As a consequence, the 3d transition metal oxides exhibit a wide range of phenomena, such as charge-transfer or Mott insulating antiferromagnets, metal-insulator transitions, transitions between paramagnetic and non-trivial magnetic phases, and charge and bond disproportionation, that originate in coupling between charge, spin and lattice degrees of freedom\cite{ZaanenPRL1985,DagottoScience2005}. This has led to intense research over the past decade to create functional materials in which transition metal oxides are cornerstones. 
How to efficiently and accurately capture correlation effects in computational modeling of materials systems has long been a crucial but elusive problem. This is important not only from a fundamental point of view, but also necessary in order to create and control functional materials in which electronic correlations play a fundamental role\cite{BESAC2015}. 
The electronic correlations in 3d transition metal compounds are often strong enough that predictive modeling using standard tools such as Density Functional Theory (DFT)~\cite{Hohenberg1964,Kohn1965} can give qualitatively and quantitatively wrong results. Titanium oxides are one class of versatile transition metal compounds that are of interest for a number of applications ranging from resistive memories to photocatalysis and dyes in paints\cite{Gratzel2000,Pechy2001,Kwon2010,Kazuhito2005}. 
Because of the many oxidation states of Ti, there exist a whole series of Ti oxides, starting with TiO$_2$, which is an insulator with a band of about 3~eV with a Ti-charge of $+4$, obtained by successively reducing the average charge of the Ti cations with concomitant removal of oxygen. An interesting subset of Ti oxides consists of the ordered Magn\'eli phases Ti$_n$O$_{(2n-1)}$. This family of compounds exhibits sharp metal-insulator transitions associated with pronounced charge and/or orbital ordering.\cite{Leonov2006} 
In particular, Ti$_4$O$_7$ is a seemingly innocuous compound with applications in resistive switching\cite{YangNanoscale2013,Kwon2010,WilliamsAdvMat2010}, oxide-based fuel cell electrodes\cite{TominakaChemComm2012}, and in catalysis\cite{YaoJMaterChem2012,WangPCCP2016}. Ti$_4$O$_7$ exhibits a complex behavior 
driven by electronic
correlations: It undergoes a phase transition between a non-magnetic conductor and semiconductor upon lowering temperature to 150~K, and a upon further reduction of temperature a transition to a magnetic semiconductor at 120~K\cite{Bartho69,Inglis1983}, and it has a surprisingly rich energy surface in its low-temperature phase, with competing ferromagnetic and antiferromagnetic states. 
 Experimental studies by Marezio {\it{et al.}}\cite{Marezio1973,Marezio84} using single-crystals in X-ray investigation on Ti$_4$O$_7$ at the transition from semiconductor-metal to semiconductor-semiconductor, and by Lakkis {\it{et al.}}\cite{Lakkis1976} combining X-ray diffraction with electron paramagnetic-resonance studies, arrived at two competing models describing the nature of the electronic structure of Ti$_4$O$_7$.  The first suggests there must be long-range order for the Ti$^{3+}$ ions and, consequently, ordered bipolarons. The second one suggested that all the bipolarons, which were well ordered in the low-temperature phase, remained in the intermediate phase, but become disordered.
  At high temperature, both models suggested the metallic behavior was due to a delocalization of the 3d electrons.\\ 
  
Because of its simple composition, being a binary transition metal oxide, and wide versatility, Ti$_4$O$_7$ is an important and representative compound serving as a testbed to elucidate the role of electronic correlations. However, despite the large number of experimental studies of 
the Ti$_4$O$_7$-phases at low, intermediate, and high temperatures\cite{Marezio1973} it has been difficult to theoretically map out these phases because of the role of the electron-electron
interactions beyond mean-field approaches such as the Local (Spin) Density Approximation (LDA)\cite{Ceperley1980} within DFT. Liborio and Harrison\cite{Liborio2008} used \textit{ab initio} thermodynamics, with electronic structure calculations within the DFT+LDA scheme, at low temperatures to predict phase stability in good agreement with experiments and also used their model to explain the formation process of the Magn\'eli phase. 
 Leonov {\it{et al.}}\cite{Leonov2006} performed an LDA+U calculation on the low temperature phase of  Ti$_4$O$_7$ and found an energy gap in good agreement with experiment and an antiferromagnetic (AF) charge-ordered ground state; Eyert {\it{et al.}}\cite{Eyert2004} performed LDA band structure calculation on the same low temperature structure but were unable to reproduce the spin-singlet ground state. More recently, Liborio {\it{et al.}}\cite{Liborio2009} have performed DFT calculations using DFT with hybrid exchange-correlation (XC) functionals (B3LYP) on low, intermediate and high temperature Ti$_4$O$_7$ suggesting an antiferromagnetic (AF) charge-ordered semiconducting state at low temperature. Weissmann and Weht\cite{Weissmannprb2011} also used LDA+U and found the same AF ground state as Leonov {\em et al.}\cite{Leonov2006} and Liborio {\em et al.}\cite{Liborio2009}\\
  
  In a more recent study, X. Zhong {\em et al.}~\cite{Zhong2015} suggested two energetically competing low-energy anti-ferromagnetic states at low temperature using an atomic self-interaction correction (SIC) scheme within DFT implemented in the SIESTA~\cite{SIESTA}
  electronic structure code, as well as more expensive hybrid functional (HSE06)~\cite{Heyd2003,Heyd2004} all-electron calculations. 
 One state was the previously studied AF state by Liborio {\em et al.}~\cite{Liborio2009}, while the other one was a new AF state with lower energy. The energy difference between these two AF states was only 4~meV per formula unit, which is small enough that one can reasonably question the validity of the conclusions.  
 
We will here present calculations of the magnetically ordered low-temperature states of the Magn\'eli-phase oxide
Ti$_4$O$_7$ using highly accurate Quantum Monte Carlo (QMC) simulations with a goal of conclusively
addressing the energy ordering of them, as well as the detailed nature of the spin ordering of the magnetic low-temperature states. We will show that QMC can be used to obtain these states and their energies, and our results also demonstrate more broadly the application of QMC to systems
with complex energy surfaces that include competing non-trivial magnetic states driven by exchange and correlation. Our QMC results confirm the results of Zhong {et al.}~\cite{Zhong2015} using LDA+SIC (and also hybrid functional calculations) regarding the ground state and adjacent low-energy states of the low-temperature phase of Ti$_4$O$_7$. This in and of itself is useful and interesting: we can validate these results obtained from the much less expensive LDA+SIC calculations. We will also analyze the results obtained from the QMC calculations
and the LDA+U and LDA+SIC calculations, as well as total energy from HSE06 hybrid functional calculations, in order to obtain some insight into where DFT calculations fail, as well as into what QMC does differently from 
LDA+U and LDA+SIC for this compound. The insight gained from this analysis will help improve DFT-based methods and guide their usage.


It has long been known that the workhorse of materials modeling, Density Functional Theory, in its standard implementations, such as the Local Density Approximation or the Generalized Gradient Approximation (GGA)~\cite{Perdew1992}, fails miserably for systems where electronic correlations are important. For example, DFT-LDA or DFT-GGA fail to predict the correct ground state for the seemingly simple binary oxide NiO\cite{Kasinathan2006}, apart from the consistently under-estimated band gap in insulators. Various extensions beyond the LDA or GGA have been devised in order to better account for electronic correlations, especially those that arise from more localized electronic orbitals, such as 3d or 4f orbitals. The LDA+U 
(or GGA+U, depending on the choice of exchange-correlation functional) 
method assigns a Hubbard-like on-site repulsion for doubly occupied orbitals and has been used rather extensively with some success to model a large range of compounds. The $U$-parameter leads to split upper and lower Hubbard bands, with a splitting that scales as $U$, which is therefore some measure of the strenght of correlations, as represented by the on-site repulsion. Nevertheless, the LDA+U~\cite{Anisimov1991a,Anisimov1991b,Anisimov1993,Anisimov1997} method suffers from being somewhat {\em ad hoc} in that it cannot be derived from more fundamental principles. Furthermore, the $U$-parameter is typically adjusted to fit something, typically a direct band gap\cite{anisimov1991}, although more recent methods\cite{Cococcioniprb2005} use self-consistency through
linear response to arrive at a value for $U$. In any case, there is no guarantee that the obtained band structure is correct, other than the band gap if that was fitted. Another method consists of removing self-interactions, as approximate DFT functionals includes interactions of each electron with itself. For itinerant electrons, self-interactions do not lead to large errors, but for localized electronic states self-interactions can lead to significant errors. Based on work by Perdew and Levy~\cite{Perdew1997}, Spaldin and Filipetti~\cite{Filippetti2003} developed a SIC scheme using a unitary transformation of localized orbitals to atomic orbitals and subsequent minimization of the self-interaction error. This scheme has been implemented in SIESTA~\cite{SIESTA}, and has shown to give good results for a number of transition metal oxides as far as the nature of the ground state and the direct band gap are concerned\cite{Svane1990,Szotek1993,Kodderitzsch2002,Fischer2009,Dane2009,Hughes2008,Archer2011}. However, this SIC scheme, while having a more \textit{ab initio} underpinning than LDA+U, still fundamentally approximates correlation energies. Furhermore, it includes an adjustable parameter $\alpha$ in the range of $[0,1]$, with $\alpha=0$ meaning no SIC at all (appropriate for metallic systems where self-interactions are not important) and $\alpha=1$ for ionic insulators, such as NaCl. For oxides such as transition metal oxides, it has been found empirically that $\alpha\approx0.5$ gives results in reasonable agreement with experiments, but there is no formal explanation for why this value is appropriate. As for LDA+U, there is no guarantee that the band structure obtained using SIC is even qualitatively correct (other than, {\em e.g.,} the band gap).

There are other methods more explicitly devised to deal with electronic correlations, but these are universally much more computationally expensive than DFT methods. The dynamical mean field theory (DMFT)~\cite{Georges1996} has been used rather extensively recently to address questions in a variety of correlated materials, such as the ``Fermi arc'' in high-T$_c$ materials\cite{Stanescuprb2006}. 
The dynamical correlations in DMFT are treated within a quantum point defect (Anderson impurity) approximation, in which the self energies depend on frequency but not wavevector, although more recent developments have extended DMFT to include some spatial variation, and therefore wavevector dependence\cite{Hauleprb2007}. 

The Density Matrix Renormalization Group (DMRG)~\cite{White1992} is in principle an exact method for obtaining the ground state of a quantum system. However, it is most suited for one-dimensional or pseudo-one-dimensional systems such as quantum spin chains or ladders. It rapidly becomes intractable as the width of the ladders increase, and it is at the present not at all applicable to real three-dimensional materials.

QMC simulations are also in principle exact and include all electron-electron interactions. The QMC method is based on a strict variational principle and so obtains a rigorous upper bound for the ground state energy by directly minimizing the expectation value of the Hamiltonian for 
a many-body wavefunction. The one fundamental approximation in QMC for fermionic systems is related to the antisymmetry of the wavefunction. The most common way to enforce antisymmetry is through the fixed-node approximation, in which the nodes of the variational wavefunction are fixed based on an initial antisymmetric trial wavefunction. Note that even with this approximation, QMC still obeys the 
variational principle and yields an upper bound to the exact ground state energy, but the quality the upper bound depends on the nodal structure of the initial trial wavefunction. QMC is usually implemented in two steps. The first step is an initial variational Monte Carlo minimization in which Jastrow-type (or dynamical) correlation factors are added to an initial Slater determinant trial wavefunction, and these correlation factors contain some adjustable parameters over which the energy is minimized. The second step is a diffusion Monte Carlo (DMC)~\cite{Foulkes2001} minimization, in which the many-body Schr\"odinger equation is written as a diffusion equation in imaginary time. The expectation value of the Hamiltonian is then minimized by propagating a convolution of the wavefunction with the Green's function in imaginary time using many random walkers. Due to its numerical expense, QMC methods were for a long time limited to model systems or small atoms or molecules. However, because the random walkers in the DMC minimization are only weakly dependent, these methods scale linearly with nearly any number of cores available (so long as the memory per core is sufficient). The arrival of leadership-class computers such as the Blue Gene/Q or Cray Titan XK7 with hundreds of thousands of cores have led to QMC methods being routinely applicable to a wide variety of systems, from simple molecules\cite{Morales2012,Benali2014,Mitas2013,kent2014,Rocca2007,Filippi2011,Guidoni2012} 
to condensed phase metals~\cite{Shulenburger2013a,Reboredo2012,Santana2015,Wagner2015,Reboredo2014,Sorella2015,Falko2015}. 
More recently, Shulenburger and co-workers\cite{ShulenburgerACS2015} applied QMC to study inter-layer interactions in bulk and few-layer phosphorous and showed that DFT-based methods including dispersive forces cannot account for a charge redistribution that arises because of inter-layer interactions. Here, we use QMC to provide an upper bound and error estimates for the LT-phases of Ti$_4$O$_7$, with 22 atoms in the unit cell and several different magnetic states, and to analyze the differences between the essentially exact QMC results and those obtained from LDA+U and LDA+SIC. As a side result, we also show how DMC can be used in an {\it un-biased} way to select a value for the Hubbard
$U$-parameter in an LDA+U scheme.

 \section{Methods}
\subsection{QMC calculations}

This study was performed using DMC within the fixed node approximation~\cite{Anderson1980}. Since QMC, and more specifically, DMC, have been thoroughly reported~\cite{Foulkes2001,Urmigar1993,bHammond1994}, we will only briefly describe the main features of our implementation in the QMCPACK package~\cite{QMCPACK1,QMCPACK2,QMCPACK3}.\\
To describe our systems we used a single-determinant Slater-Jastrow wave function $\Psi\left(\bf{x_1,...,x_N}\right)$ expressed as 
\begin{equation}
\label{WaveFunc}
\Psi\left(\bf{x_1,...,x_N}\right)=e^{J\left(\bf{x_1,...,x_N}\right)}\Psi_{AS}\left(\bf{x_1,...,x_N}\right)
\end{equation}
where $\bf{x_i} \equiv \verb+{+ \bf{r_i,\sigma_i}\verb+}+$ is a space-spin coordinate, $J\left(\bf{x_1,...,x_N}\right)$ is the Jastrow factor describing the correlation between electrons, $\Psi_{AS}(\bf{x_1,...,x_N})$ is the antisymmetric fermionic  part of the wavefunction and N is the total number of electrons.\\

The antisymmetric part is calculated within the framework of the DFT, using the plane wave pseudopotential code PWSCF contained in the Quantum ESPRESSO (QE) simulation package~\cite{Q-Espresso}. We used the rotationally invariant LDA+U implementation in QE with the LDA XC functional proposed by Perdew and Zunger~\cite{Perdew1981} and a single Slater coefficient $F^0$ for the $U$-parameter. This is used as an additional variational optimization parameter. Convergence with respect to the kinetic energy cutoff ($E_{cut}^{wfc}$), the density cutoff ($E_{cut}^{rho}$) and size of Monkhorst-Pack~\cite{Monkhorst1976} mesh of \textit{k}-points were checked in order to achieve the same precision in all calculations, leading to the following values: $E_{cut}^{wfc}=300$~Ry, $E_{cut}^{rho}=1200$~Ry and a ($17\times 17\times 17$) \textit{k}-point grid in the HCP primitive cell for the Ti bulk calculations. The single-atom calculation calculations were performed in cubic boxes with 30~a.u dimension ensuring an insignificant amount of interaction between atoms and their periodic 
images.\\

The Jastrow factor can be factored to one-body (effective electron-ion interactions), two-body (electron-electron interactions), three-body (electron-electron-ion interactions) and higher-body orders $(J=J_1J_2J_3..)$. In this work, using one-body and two-body terms provided a satisfactory trade-off between computational expense and ease of optimization:

\begin{equation}
\label{Jastrow}
J(\bf{r_1,...,r_N})=\sum_{iI}\chi_I(\bf{r}_{iI})-\sum_{i<j}u(\bf{r}_{ij}),
\end{equation}

where $\chi$ and $u$ are one-dimensional optimized splines;  $i$ labels the electron position, $I$ labels the ion position, $r_{ij}=|\bf{r_i}-\bf{r_j}|$ is the electron-electron distance and $r_{iI}=|\bf{r_i}-\bf{R_I}|$ is the electron-ion distance. In this definition, $\chi$ is the cusp-less electron-Ion Jastrow factor, while $u$ describes the electron-electron correlations. Different correlation factors were used for like and unlike spins.  All the Jastrow parameters were optimized within the energy minimization framework~\cite{Umrigar2007} by minimizing a combination of the total energy and variance.  Since the Jastrow prefactors are strictly positive, the nodes of the Slater-Jastrow wave function are determined by $\psi_{AS}$ and therefore the fixed-node error in the DMC calculations is entirely due to the Slater determinant obtained from the DFT calculation.\\   

All calculations were performed within the pseudopotential (PP) approximation, in which the core electrons are replaced by a nonlocal potential operator~\cite{Foulkes2001}. In the DMC part, we used the locality approximation~\cite{mitas1991} to evaluate the nonlocal PP.  Norm-conserving pseudopotentials~\cite{Hamann1979} (NCPP) were used for both QMC and DFT calculations. They were defined with $(3s^23p^63d^24s^2)$ valence electrons for Titanium atoms and $(2s^22p^4)$ for Oxygen atoms and generated with the OPIUM~\cite{OPIUM} code. The oxygen PP was generated and optimized by Shulenburger and Mattsson\cite{Shulenburger2013a}. The Ti PP was generated by Krogel {\it{et al.}}\cite{Reboredo2016} using the Opium code following Rappe {\it{et al.}} scheme\cite{Rappe1990}. The many-body wavefunction was evaluated using real space cubic b-splines~\cite{Williamson2001} for speed, and we used a minimum of 8192 walkers with a 0.005 a.u time step for the DMC calculations with extrapolations to the zero timestep limit.

\subsection{Finite-size effects - Solids}
When determining properties of solids in the thermodynamic limit with QMC using periodic boundary conditions, it is necessary to control finite-size errors that arise from the spurious correlations created by the periodic images of the system.  These errors can be grouped into two categories: single-particle effects and many-body effects.
%
%
In single-particle theories (and effective single-particle theories such as DFT), calculations are usually reduced to the primitive unit cell. Then the properties of a system in the thermodynamic limit may be determined by using periodic boundary conditions and integrating over the first Brillouin zone.  Although such an integration is also used in a many-body theory such as DMC, there is an additional finite-size error due to image cell correlations, and more sophisticated techniques are needed to extrapolate to the thermodynamic limit.  For this reason, DMC calculations of periodic solids are generally performed using supercells to minimize this error. 
When simulating systems with periodic boundary conditions, the energy contributions from the classical Coulomb interactions cannot be obtained by the sums over the infinite number of images do not converge when using the classical $1/r$ Coulomb potential. The common solution is to use the Ewald summation techniques~\cite{Ewald1921}. However, the periodicity of the system implies the periodicity of the electronic correlations, which introduces effects of electrons interacting with their periodic images in neighboring cells. This unphysical error corresponds to an XC energy of a system with a periodic XC hole. These errors are not present in DFT simulations because the XC interactions are derived from the electron gas converged at the limit of infinitely large simulation cells. The first option for finite-size correction consists of increasing the system size in order to reduce the spurious effective interactions between an electron and its XC hole. 
This leads to a rapidly increasing cost of the calculations with the number of electrons in the simulation cell. The second finite-size correction is to add the model periodic Coulomb (MPC) interaction to the Ewald energy as suggested by Fraser {\it et al}~\cite{Fraser1996}.  A final correction is applied to the kinetic energy following the scheme of Chiesa~\cite{chiesa2006} {\it et al}. For all calculations in this study, energy convergence was reached at simulation cells containing 88 atoms and applying both MPC and Chiesa corrections.

\begin{figure}
 \includegraphics[angle=0,scale=0.5]{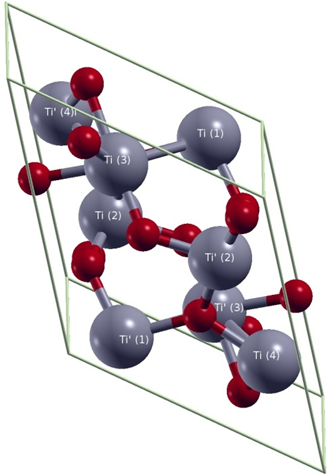}
 \caption{\label{fig:Ti4O7_cell}
Simulation cell of (Ti$_4$O$_7$)$_2$ containing 22 atoms (14 oxygen and 8 titanium)}
 \end{figure}

In periodic boundary conditions, the Hamiltonian is invariant with respect to translating any particle around the periodic boundaries. According to Bloch’s theorem, this implies that that any solution can be characterized by a given vector twist angle ${\mathbf \theta}=(\theta_x,\theta_y,\theta_z)$, defined by
\begin{equation}
\Psi({\mathbf r}_1+L_x\hat x,{\mathbf r}_2,\ldots,{\mathbf r}_N)=e^{i\theta_x}\Psi({\mathbf r}_1,{\mathbf r}_2,\ldots,{\mathbf r}_N),
\end{equation}
with similar definitions for $\theta_y$ and $\theta_z$.
In order to obtain results in the thermodynamic limit, all $k$-points in the Brillouin zone have to be averaged over. Because a change in the twists changes the grid over which the kinetic energy is calculated, an average over twist angles is an efficient way to approach the thermodynamic limit. 
Specifically, an average is taken of expectation values over all the $k$-points in the primitive cell\cite{Ceperley2001}.
DMC calculations at different twist angles are statistically independent, so the cost of twist averaging is negligible except for initialization and equilibration. The error can therefore be reduced by increasing the density of the twist grid. 
To take into account the two-body-finite-size effect, we increased the size of the supercell until the change in energy from one supercell size to the next is equal to 0.1mHa/unit-cell, leading to a supercell 4 times larger than the primitive cell (Figure-\ref{fig:Ti4O7_cell}) and 32 boundary twist per supercell. 
 
Ti$_4$O$_7$ at low temperature (less than 120~K) was modeled using two formula unit cells (Ti$_8$O$_{14}$) to account for the various spin configurations, as shown in Fig.~\ref{fig:Ti4O7_cell}.
The enlarged unit cell thus contains one chemical unit cell plus one image of it. In order to distinguish between four Ti atoms and their images in the enlarged unit cell, we adopt the following notation for the remaining of the discussion; (Ti$1$,Ti$2$, Ti$3$, Ti$4$) and their images (Ti1', Ti2', Ti3', Ti4'). The atomic spins around the Titanium atoms Ti$2$, Ti$4$, and their periodic images Ti2', Ti4', in the simulation cell  are very small. 
 Zhong {\it et al}\cite{Zhong2015}, using LDA+SIC 
 as well as the HSE06 hybrid functional, identified a ferromagnetic state (FM), where all Ti atom have a spin up ($\uparrow$),  and three anti-ferromagnetic state, AF1 AF2 and AF3, as shown in Table~\ref{tab:Spin}.
\begin{table}[t]
\caption{\label{tab:Spin}
Spin configuration for Titanium atoms in Ti$_8$O$_14$ Ferromagnetic State (FM) and AntiFerromagnetic states (AF1) (AF2) and (AF3). ($\uparrow$) represents a spin up while ($\downarrow$) represents a spin down.  The spins are for all Ti$_x$ atoms and their images Tix' in the sublattice as follow (Ti$1$ , Ti$2$, Ti$3$, Ti$4$ | Ti1' ,Ti2' ,Ti3' ,Ti4'); 0 denotes an almost null atomic spin. 
}
 \begin{tabular}{cc}
\hline
 &    Spin Configuration  \\
\hline
FM & ($\uparrow,\uparrow,\uparrow,\uparrow | \uparrow,\uparrow,\uparrow,\uparrow$) \\
AF1 & ($\uparrow,0,\downarrow,0\, | \uparrow,0,\downarrow,0$)\\                                       
  AF2 &   ($\uparrow,0,\uparrow,0\, | \downarrow,0,\downarrow,0$)  \\
AF3  &   ($\uparrow,0,\downarrow,0\, | \downarrow,0,\uparrow,0$)  \\
\hline
  \end{tabular}
\end{table}
%
Because the cohesive energy of AF2 was previously found to be clearly higher than the two other states, we decided not to consider it in this study and will focus on FM, AF1 and AF3 states.\\

\section{Results and Discussion}
\begin{figure}
 \includegraphics[angle=-90,scale=0.4]{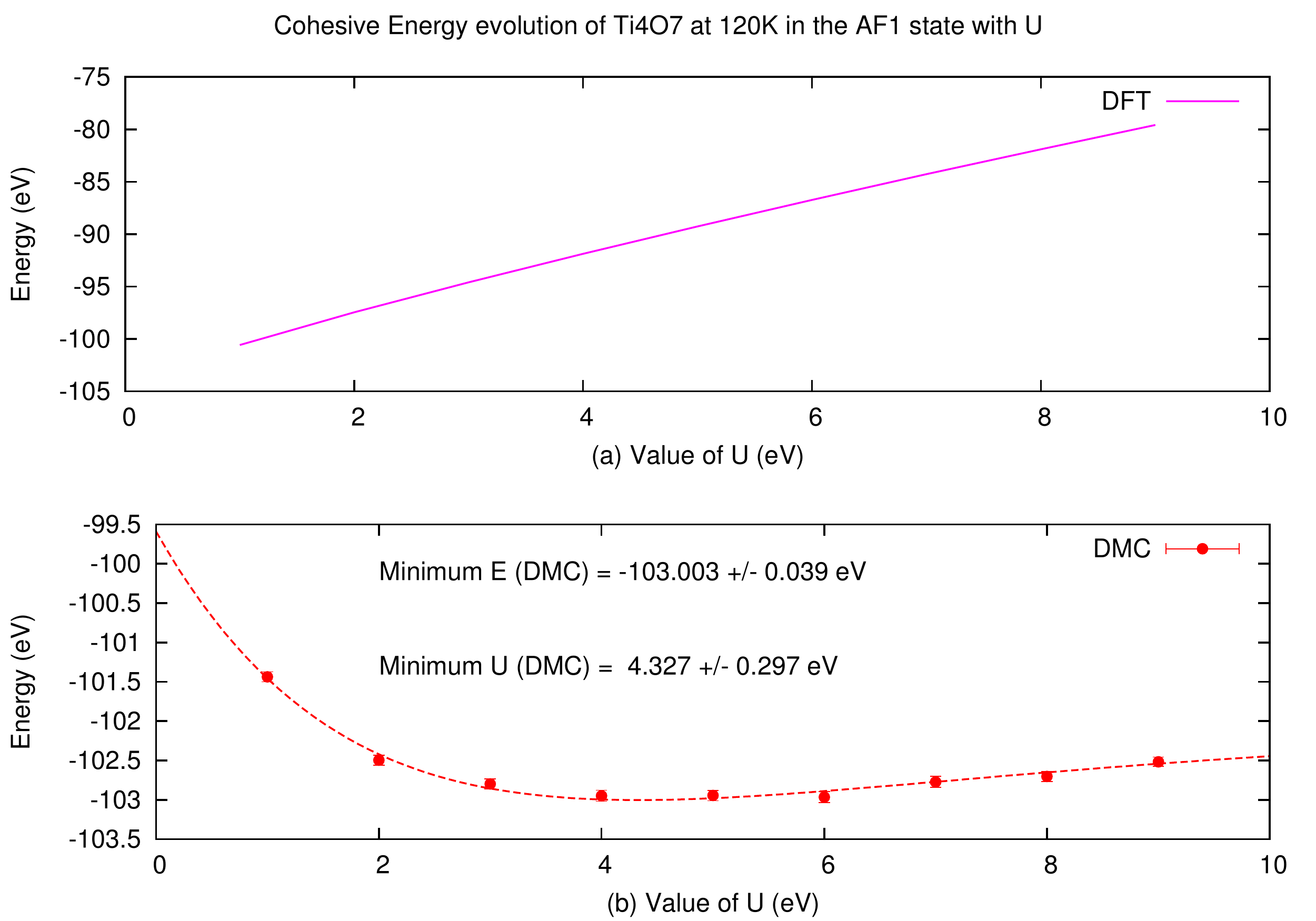}
 \caption{\label{fig:Ti4O7_U}
DFT (a)
and DMC (b) energies of Ti$_4$O$_7$ as a function of the value of $U$ for the AF1 state. The DFT curves increases linearly with the value of U while 
DMC one is  non linear function showing a minimum in energy at 
$U=4.3$~eV for DMC.}
 \end{figure}

As we discussed in previous sections, the results obtained using DFT are highly dependent on the choice of the XC functionals. In the case of the antiferromagnetic states of Ti$_4$O$_7$, none of the tested functionals (LDA or GGA)  was able to reproduce or predict the AF states (they were not obtained as stable states). Using LDA+U, LDA+SIC, or the HSE06 hybrid functional, the AF states are stable but the difference in energy is only 4 - 5~meV per formula unit. In previous studies using  LDA+U\cite{Leonov2006,Eyert2004}, the value of $U$ was set to reproduce the experimental band gap of Ti$_4$O$_7$ ($U = 3.0$~eV and an exchange coupling of $J=0.8$~eV), and Weismann and Weht\cite{Weissmannprb2011} used a value of $U=5.44$~eV obtained from considering energy differences between phases and obtained a band gap of 0.7~eV. In our LDA+U calculations, we used LDA+U as implemented in the Quantum Espresso code with a
single Slater coefficient $F^0=U$. To select the value of $U$ in our study, we chose to use the variational nature of QMC to scan through the values $U$. Doing so, we remove the ad-hoc empirical contribution to the Hamiltonian. As expected and shown on Fig.~\ref{fig:Ti4O7_U}(a), increasing the value of $U$ from 1 to 10, which is equivalent to populating the d band, leads to a linear increase of the energy at the DFT level.
In contrast, the total DMC energy (as well as the VMC one) of the AF1 state as a function of the trial wavefunction, characterized by the $U$-parameter, follows a nonlinear U dependence with a minimum at 
$U_{DMC}=4.3 \pm 0.3$~eV [Fig~\ref{fig:Ti4O7_U} (b)]. 
The
minimum energy of DMC is limited only by the nodal surface, derived from the $U$-dependent trial wavefunction.
This shows clearly the sensitivity of the energetics to the $U$ parameter as more sophisticated treatments of dynamic correlation are used; whereas the variation in  the value of $U$ between 1~eV and 10~eV leads to a spread of 20~eV in LDA+U, it only leads to a spread of 
1.5~eV in DMC. In the remaining of the calculation we used the $U_{DMC}=4.3$~eV value for all the studied systems; this value is smaller than the value of 5.4~eV used by Weissmann and Weht\cite{Weissmannprb2011}. It is interesting to note that this unique method selection of $U$ is quite 
different from more recent LDA+U methodologies, 
where $U$ is selected within a 
rotationally invariant simplified LDA+U scheme (keeping only the $F^0$ Slater
coefficient) by enforcing 
self-consistency between the chosen $U$ and the value of $U$ obtained from linear response. Within
that scheme, the total energy is still a monotonic function of $U$, in contrast with the dependence of the energy on 
$U$ obtained here from DMC. It should be pointed out that there does not exist, to the best of our knowledge, any
fundamental justification for either method of selecting $U$.

Table~\ref{tab:Energy_Diff} shows the results of our DMC calculations of the energies of the 
FM, AF1, and AF3 states, as well as the energies from LDA+U,  LDA+SIC\cite{Zhong2015} using $\alpha=0.5$, and also from HSE06 hybrid functional\cite{Zhong2015};
Table~\ref{tab:E_Gap} shows the energy gaps obtained from
LDA+U, LDA+SIC calculations ($\alpha=0.5$), and HSE06 hybrid functional. 
We note that neither the $U$-parameter in LDA+U or the $\alpha$-parameter in LDA+SIC in our calculations was adjusted to fit the energy gaps. The LDA+U, LDA+SIC, and HSE06 hybrid functional calculations 
show a clear separation in energies between the FM state and the AF states, especially the LDA+U and HSE06 calculations, while the energy difference between AF1 and AF3 is very small (4 - 5~meV), clearly beyond errors within LDA+U, and LDA+SIC. However, the DMC results increase significantly the energy differences between the AF states, favoring the AF3 state (staggered antiferromagnetic order within each sublattice and between sublattices) over both the AF1 (staggered antiferromagnetic order within each sublattice, but ferromagnetic order between sublattices) and the FM states, although the DMC energy separation is substantially smaller between the FM and AF3 state than obtained from LDA+U. Note that the DMC calculations also have error bars and the energy differences between the states
are significantly larger than the error bars. It is important to note that this ordering is identical to what is found with DFT; the main difference is the accuracy of the prediction. The energy difference found between FM and AF3 at the DMC level of theory becomes much closer to the LDA+SIC result despite starting from a much larger difference in the LDA+U calculations which superficially suggests a somehow a better correction to the XC interactions for LDA+SIC than for LDA+U. We will argue later that this is not necessarily the case.

We have also calculated the energy differences between the AF3 and AF1 states for a range of values of the $U$-parameter in LDA+U, for the $\alpha$-parameter in LDA+SIC, and for the exchange mixing parameter $a$ in the HSE hybrid functional. We note that recent work shows that the choices of the exchange mixing of $a=0.303$ and range $\omega=0.201$~{\AA}$^{-1}$ specified in HSE06 are optimal\cite{Moussa-2012}. The energy difference between AF3 and AF1 is relatively insensitive to the values of $U$,  $\alpha$, or exchange mixing $a$. As $U$, $\alpha$, or $a$ increase, the 3d states become more localized, and a reduced overlap leads to a smaller energy difference. At the same time the band gap increases by pushing apart the split bands in LDA+U, or by reducing the energy of the occupied orbitals relative to the unoccupied ones in LDA+SIC, or by increasing the exchange interactions between 3d states as $a$ increases. When $U$ varies from 1~eV to 7~eV, the energy difference from LDA+U only varies from 4.1~meV to 1.6~meV. Similarly, when $\alpha$ varies from 0.3 to 0.7, the energy difference varies from 4.5~meV to 2.3~meV, and when $a$ varies from 0.15 to 0.4, the energy difference decreases from 5.3~eV to 3.3~meV. On the other hand, the band gaps increase approximately linearly from about 0.3~eV to about 2.8~eV for LDA+U, from about 0.2~eV to 1.7~eV for LDA+SIC, and from about 0.5~eV to 3~eV for the HSE functional. Moreover, as U, $\alpha$, or $a$ decrease, the local moments decrease because of delocalization of the orbitals. For example, the (Mulliken population) moments from LDA+SIC decrease in magnitude from 0.93 to 0.68 as $\alpha$ decreases from 0.7 to 0.3, and the orbital moments in HSE decrease from 0.96 to 0.76 as $a$ decreases from 0.4 to 0.15: improving the band gap by decreasing $\alpha$ or $a$ makes the magnetic moments worse while the energy difference between AF3 and AF1 barely changes.\\        
\begin{figure}
 \includegraphics[angle=0,scale=0.2]{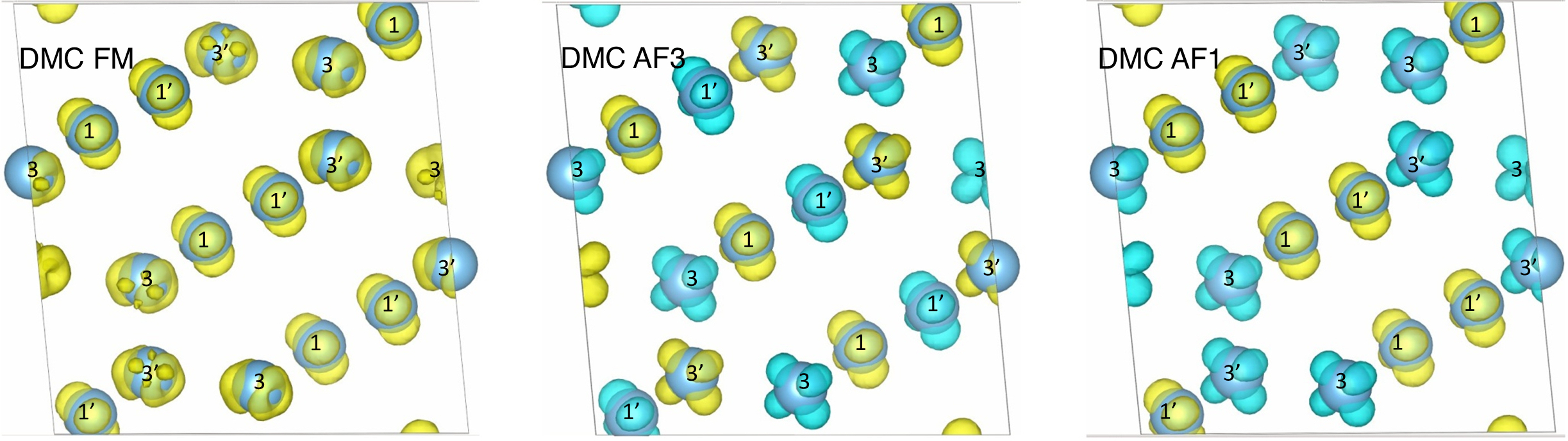}
\includegraphics[angle=0,scale=0.2]{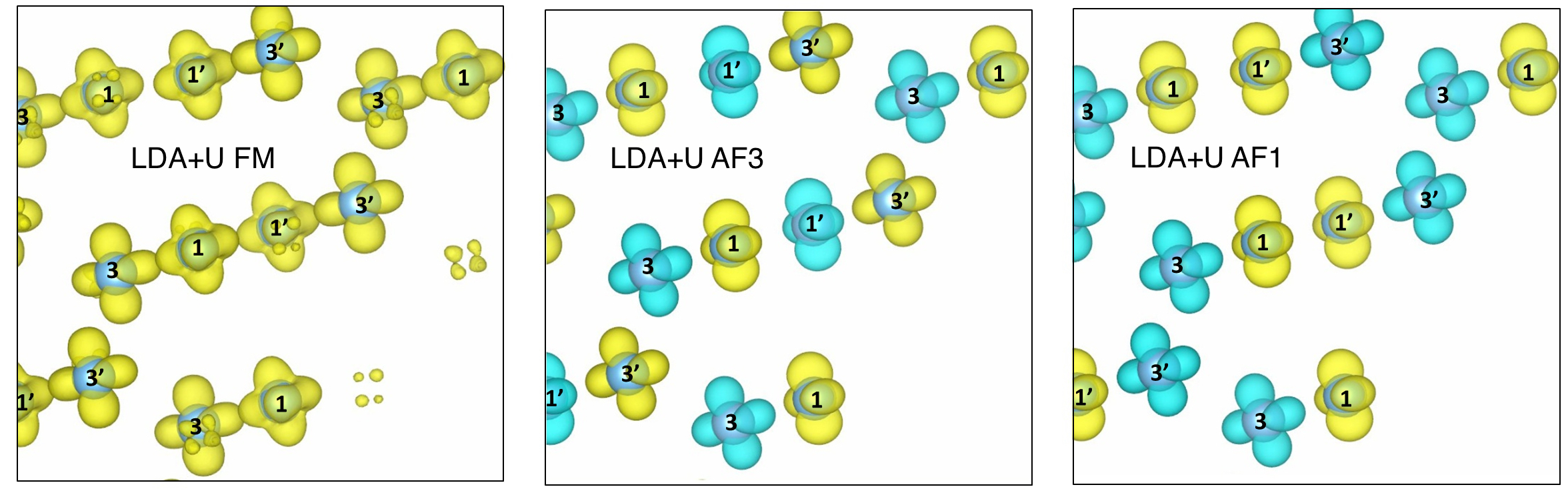}
 \includegraphics[angle=0,scale=0.5]{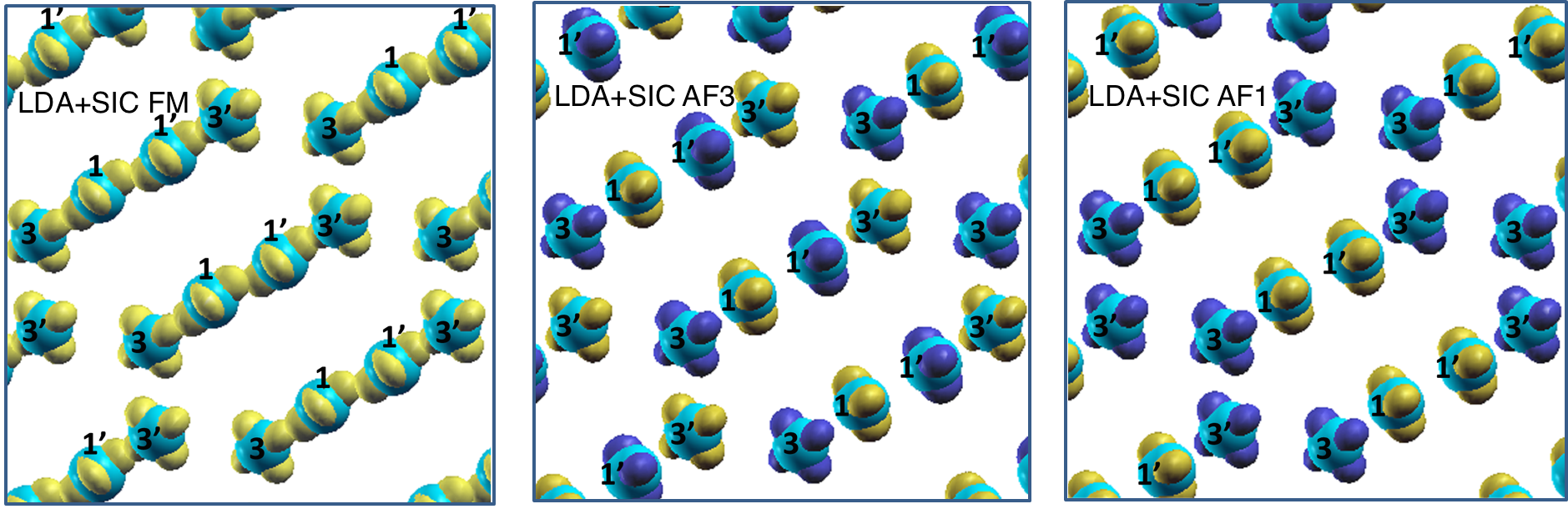}
 \caption{\label{fig:AF-Spin}
 Spin density difference ($\rho_\uparrow-\rho_\downarrow$) from the DMC (top), LDA+U (center) and LDA+SIC (bottom) depicting iso-surfaces of 0.0298 electrons per $a_0^3$, where $a_0$ is the Bohr radius, in the FM state (left panel), the AF3 state (middle panel), and the AF1 state (right panel). The value of the iso-surfaces was chosen to give a good visual representation of the spin densities around the Ti$^{3+}$ sites. 
Yellow represents up-spin ($\rho_\uparrow-\rho_\downarrow=0.0298/a_0^3$), blue down-spin	
($\rho_\uparrow-\rho_\downarrow=-0.0298/a_0^3$)
}
 \end{figure}
 
 
 \begin{table}[t]
\caption{\label{tab:Energy_Diff}
Energy difference between the antiferromagnetic state (AF3) and the ferromagnetic state (FM), and the antiferromagnetic state (AF2) and the ferromagnetic state (FM) in eV per formula unit using LDA+U, LDA+SIC and DMC. A positive value means that the AF3 state is lower in energy. 
}
 \begin{tabular}{llccllll}
\hline
 Method &    E(FM-AF3)      & E(AF1-AF3) \\
\hline 
LDA + U & +0.608 & +0.004 \\
LDA + SIC\cite{Zhong2015} & +0.118& +0.004\\
HSE06 & +0.605 & +0.005\\
\hline                                                    
  DMC  &   +0.17(1)  & +0.07(1)\\
\hline
  \end{tabular}
\end{table}

 \begin{table}[t]
\caption{\label{tab:E_Gap}
Energy gaps in eV obtained from LDA+U, and LDA+SIC calculations for the FM, AF1, and AF3 states. The experimental gap is about 0.25~eV.
}
 \begin{tabular}{llccllll}
\hline
 Method &    FM      & AF1 & AF3 \\
LDA + U & 1.32 & 1.55 & 1.61\\
LDA+U\cite{Weissmannprb2011} & 0.53 & 0.75 & N/A\\
LDA + SIC\cite{Zhong2015} & 0.61& 0.84 & 0.94\\
HSE06 & 0 & 1.36 & 1.47\\
\hline
  \end{tabular}
\end{table}
Figure \ref{fig:AF-Spin} shows the DMC iso-surfaces of the spin densities 
with yellow representing net up-spin (positive spin density) and blue representing down-spin (negative spin density). We note first of all the spin moments on Ti1 and Ti1' have essentially the same $d_{yz}$ spatial distribution (with the x-axis the projected line connecting Ti3 - Ti1 - Ti1'- Ti3') in all three magnetic states -- the moments only change sign between the different magnetic states. There is only a very slight rotation of the moment distributions on Ti3 and Ti3' in the AF state that make them appear like a linear combination of $d_{yz}$ and $d_{x^2-y^2}$ orbitals.
In contrast, the spin distributions on 3 and 3' have different shapes in the FM state compared to in the AF states and are neither pure $d_{z^2}$ or four-lobed planar $d$-orbitals. 

For comparison, the middle row of Fig.~\ref{fig:AF-Spin} depicts iso-spin density surfaces obtained from LDA+U. This figure shows that in LDA+U too the shapes of the spin density distributions in AF3 (center) and AF1 (right) are identical, and extremely similar to those obtained from DMC, with $d_{yz}$ character on Ti1 and Ti1', and a combination of $d_{yz}$ and $d_{x^2-y^2}$ on Ti3 and Ti3', also in good qualitative agreement with the results of Leonov {\em et al.}\cite{Leonov2006}. In contrast, the spin distributions in the FM state show some marked differences both from those in the AF states as well as those in the DMC FM state. {\em All} spin distributions in FM have approximately the same $d_{yz}+d_{x^2-y^2}$ character, with spin density lobes much more directed along the line connecting the Ti3 - Ti1 - Ti1' - Ti3' sites. This clearly leads to a different near-neighbor overlap in the LDA+U FM than in the DMC FM state.

The lower panels of Fig.~\ref{fig:AF-Spin} show iso-surfaces of the spin densities obtained from the LDA+SIC calculations. The spin distributions in AF3 and AF1 on Ti1 and Ti1' again agree very well with the DMC and LDA+U results. The shapes of the distributions at Ti1 and Ti1' and also at Ti3 and Ti3' are completely different in the FM state between DMC and LDA+SIC. Now the Ti1 and Ti1' sites have much more $d_{xz}$ character, while the Ti3 and Ti3' sites have $d_{yz}+d_{x^2-y^2}$ character. Again, the differences in spin distributions obtained from LDA+SIC compared to DMC will obviously lead to different overlaps between near neighbor sites.

Before we discuss these results in detail, it is worth recapitulating the basic ingredients in 
LDA+U and LDA+SIC. First, both these methods as usually implemented involve projecting
the single-particle states on a local atomic basis. In LDA+U, a Hubbard-like energy that penalizes two electrons (spin-up and spin-down) on the same atomic orbital is added to the
total energy. Another term that (approximately) corrects for double-counting also has to be added to the total energy. The effective potential added to the Kohn-Sham equations then contains a spin-independent and a spin-dependent part, in addition to the standard
LDA or GGA potentials. Note that this construct projects out self-interactions between
the different angular momentum atomic orbitals. The spin-dependent part, together with the LDA or GGA potentials,
give rise to an effective exchange interaction between neighboring sites\cite{Liechtensteinprb1995}. This interaction comes from a {\em local} spin-dependent potential that arises from the LDA approximation and the on-site $U$-interaction. Therefore,
the near-neighbor exchange is a consequence of the single-particle states not being perfectly localized on each atomic site, but having finite overlaps with atomic orbitals on
neighboring sites. Nevertheless, LDA+U has been successful in obtaining a reasonable ground state for a number of antiferromagnetic Mott insulators or charge transfer insulators, in addition to obtaining reasonably accurate values for spin moments and effective exchange couplings\cite{Shickprb1999,Anisimov1997,Leeprl2005}. 

The SIC implementation\cite{Pemmarajuprb2007} also projects the single-particle states onto localized orbitals by
seeking a unitary transformation to localized orbitals that minimizes the self-interaction energy between these orbitals. Main differences between SIC and LDA+U are that SIC only operates on occupied orbitals and pushes them down in energy, while the unoccupied orbitals are not affected, and that SIC does not add additional {\em ad hoc} interactions
to the Hamiltonian; SIC only tries to eliminate self-interactions. Both LDA+U and SIC correct the tendency of local or semi-local XC
functionals to erroneously delocalize d-orbitals of transition metals. This delocalization, which is largely due to incorrect self-interactions of the orbitals, leads to
overestimating the Heisenberg coupling between local moments. Interestingly, the self-interaction free Hartree-Fock approximation tends
to over-correct and leads to too small effective Heisenberg couplings, presumably
because of a lack of any correlation energy. An effective Heisenberg interaction arises in SIC, as in LDA+U, from the overlap of atomic orbitals on neighboring sites, but the strength of the interaction now comes only from projecting out self-interactions (in addition to the local LDA xc-contribution) and has no contribution
from an added local interaction, as in LDA+U. The SIC implementation in SIESTA has also been used to calculate spin moments as well as effective Heisenberg exchange in a number of structures, but the results for the Heisenberg couplings are mixed with relatively good results for structures with half-filled d-bands but worse away from half filling\cite{Akandejpc2007}. Notably, away from half filling, SIC appears to need different values of the
$\alpha$-parameter depending on which quantity
is sought, the band gap or the exchange couplings. 

\begin{figure}
\includegraphics[scale=0.5]{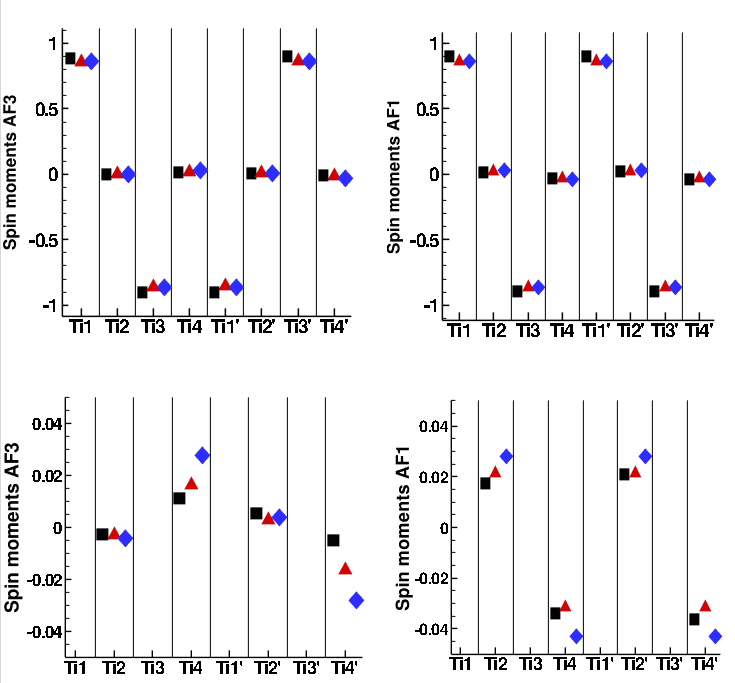}
 \caption{\label{fig:spin_moments}
Calculated spin moments from DMC (black squares), LDA+U (red triangles), and LDA-SIC (blue diamonds) for AF3 (left panels) and AF1 (right panels). The bottom row shows the moments on an enlarged scale. The moments were obtained by integrating spheres of radius 1.4~{\AA} centered on the Ti atoms.}
 \end{figure}

\begin{figure}
\includegraphics[scale=0.5]{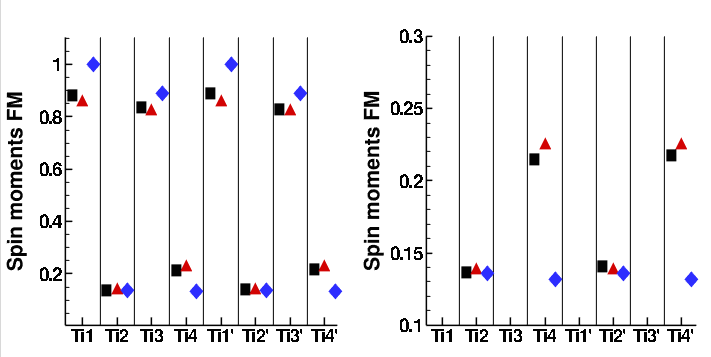}
 \caption{\label{fig:spin_moments_FM}
Calculated spin moments from DMC (black squares), LDA+U (red triangles), and LDA-SIC (blue diamonds) in the FM state in units of the Bohr magneton. The right panel shows the moments on an enlarged scale. The moments were obtained by integrating spheres of radius 1.4~{\AA} centered on the Ti atoms.}
 \end{figure}

With this background we now turn to discussing the results in more detail. We first calculate the local spin moments associated with each Ti atom. The idea is that the energy differences between the states can be analyzed using a Heisenberg model\cite{Zhong2015}, and this analysis can therefore also give some insight into how LDA+U and LDA+SIC treat electron-electron interactions differently and incorrectly. Figures~\ref{fig:spin_moments} and \ref{fig:spin_moments_FM} shows the calculated spin moments obtained by integrating the spin density in a sphere of radius ranging of 1.4~{\AA} centered on the Ti atoms in the unit cell. Note that the spin moments on Ti1, Ti1', Ti3, and Ti3' are very nearly equal in magnitudes in the FM, AF1, and AF3 states, while the spin moments on the remaining Ti ions are close to zero ($\sim0.01$). 
 The smallest Ti-Ti distances between the large-moment ions
are $r_1=2.802$~{\AA}, $r_2=3.133$~{\AA}, and $r_3=3.159$~{\AA} between 
Ti1 and Ti3 (and Ti1' and Ti3'), Ti1 and Ti1' (and Ti3 and Ti3'), and Ti1' and Ti3 (and
Ti1 and Ti3'). This lends justification to the hypothesis that the energy differences are to lowest order driven by nearest-neighbor exchange coupling. First of all, the large spin moments on Ti1, Ti3, Ti1', and Ti3' are nearly identical in magnitude both for DMC, LDA+U, and LDA-SIC in AF3, in FM, as well as in AF1; the FM moments from LDA+SIC are only slightly larger in the FM state.
This is in and of itself rather remarkable: both LDA+U and SIC seem to handle charge
disproportionation and local spin moments very well. 
LDA+U captures very accurately  the 
large spin moments on Ti1, Ti1', Ti3, and Ti3' in all three phases. It also does a good job with the small spin moments on Ti2, Ti2', Ti4, and Ti4' in the FM and AF1 states; the main errors are the Ti4 and Ti4' moments in the AF3 state, with approximately twice as large
magnitude for the spin moments as obtained from DMC. LDA+SIC does a little worse in
terms of the spin moments, with the large moments slightly overestimated in the FM state, and the small Ti4 and Ti4' moment magnitudes much overestimated in the AF3 state (about 0.03
Bohr magnetons instead of 0.005 - 0.01), and slightly overestimated magnitudes of the 
small moments Ti2, Ti2', Ti4 and Ti4' in the AF1 state. For comparison, Weissman and Weht\cite{Weissmannprb2011} obtained a spin moment of magnitude 0.64~$\mu_B$ on the large-moment Ti sites, and Leonov {\em et al.}\cite{Leonov2006} obtained 0.67~$\mu_B$, and a magnitude of between 0.02 and 0.04 $\mu_B$ for the small-moment Ti sites.

We first focus on the FM and AF1 states that are separated by 0.1~eV in the DMC, compared to 0.604~eV in LDA+U, and 0.114~eV in LDA+SIC. Because of the identical spin distributions in DMC on Ti1 and Ti1' in the FM and AF1 states, the energy difference between these states must be driven mainly by an effective antiferromagnetic exchange coupling between the pairs nearest-neighbor pairs Ti1-Ti3 and Ti1'-Ti3'. Note now the small but clearly visible difference in the spin distributions on the Ti3 and Ti3' sites between the FM and AF1 states in both the DMC results and the LDA+U (Fig.~\ref{fig:AF-Spin} middle panels). The orientation of the lobes of the spin distributions in the LDA+U FM clearly suggests a larger overlap between nearest neighbors Ti1 - Ti3 and Ti1' - Ti3' than in DMC and a concomitant overestimated exchange coupling between these sites, consistent with a much too large energy difference between AF1 and FM in LDA+U, compared to DMC. The picture In LDA+SIC is less clear with the near-perpendicular orientation of the lobes on Ti1 and Ti3, and Ti1' and Ti3' in FM, compared to DMC but visual inspection of Fig.~\ref{fig:AF-Spin}, lower left panel, makes it plausible that the overlaps between the Ti1 and Ti3 sites, and the Ti1' and Ti3' sites are slightly larger in LDA+SIC than DMC, with a slightly larger exchange coupling and energy difference FM-AF1.  


LDA+U severely overestimates the Ti1 - Ti3, and the Ti1' - Ti3' exchange coupling, while LDA+SIC slightly overestimates it. 
Increasing $U$ more strongly localizes the 3d-like electronic states on single sites and reduces
overlap with adjacent sites, which would reduce the effective Heisenberg exchange and therefore reduce the energy difference between the FM and AF3 states (neglecting the change in kinetic energy because of reduced near-neighbor hopping). It would also reduce the energy gap, which is significantly overestimated with this value of $U$. But then the atomic
moments (and charges) would be overestimated. This points to a problem, at least with
the implementation of LDA+U used here: the value of $U$ is ``right" for the localized moments but gives rise to too strong near-neighbor exchange and a much too large energy gap. Furthermore, the shape and orientations of the spin distributions in the FM state are not right in LDA+U, and no adjusting of the $U$-parameter can correct that. 

In contrast, LDA+SIC appears to localize the moments slightly
too much in the FM state, with moments about 10\% too large. 
This is reminiscent of the general trend of the Hartree-Fock approximation, where exchange-only leads to too strong near-neighbor same-spin repulsion and orbital localization; some corrections for electronic correlations are needed. However, the energy gap from LDA+SIC is also too large (see Table~\ref{tab:E_Gap}). Decreasing $\alpha$ will reduce the localization of the 3d-orbitals, which increases the near-neighbor Heisenberg exchange and reduces the energy gap\cite{Zhong2015}. If we use $\alpha=0.3$ instead, of $\alpha=0.5$,  the energy gap is reduced to about 0.2~eV for the AF1 and AF3 states, but now the FM state is metallic. The energy differences become 9~meV between AF1 and AF3, and 128~meV between FM and AF3; the energy differences increase somewhat as the moments delocalize more with reduced self-interaction corrections. The problem is just that the moments have delocalized too much with and almost 20\% reduction of the large moments in the AF1 and AF3 states and a somewhat smaller reduction in the FM state. Also, the shape and orientations of the spin distributions are not right in the LDA+SIC FM state, and tuning the $\alpha$-parameter is not going to correct the spin distributions. 

We note that HSE06 calculations, similarly to LDA+U, gives too large an energy separation between the FM and AF3 states while the energy gap closes in the FM state. This is indicative of too short range of the exchange interaction, with the consequence that the orbitals are too delocalized in the FM state.

The differences between the AF3 and AF1 states are much more subtle. First of all, the energy difference is much smaller than between the AF1 and FM states, only about half from DMC, while LDA+U and LDA+SIC severely underestimates this energy difference at only 1~\%-3~\% of the AF3-FM one. Within a Heisenberg picture, the energy difference between these two states is driven by next-nearest neighbor exchange between the pairs Ti1 and Ti1', and Ti3 and Ti3'. This picture is supported by the fact that the shapes of the moment distributions are identical (see Fig.~\ref{fig:AF-Spin}). However, the large atomic moments are very nearly equal for QMC, LDA+U, and LDA+SIC. Therefore, the energy difference must be driven
either by underestimated coupling between these pairs of moments, or by differences in 
other pairs of interactions, such as between Ti1 and the small moment Ti4, and between Ti1' and Ti4'. The inter-atomic distance between these pairs of atoms is also 3.133~{\AA}, the same as between Ti1 and Ti1'. Because of this, and because of the very much smaller moments in Ti4 and Ti4', it seems unlikely that the these interactions are relevant: one would expect the overlaps between the Ti1 and Ti1' orbitals to be approximately the same as between the Ti1 and Ti4 atoms so the smaller moments on Ti4 and Ti4' would make the interaction between Ti1 and Ti4 about 1\% of the interaction of the exchange interaction
between Ti1 and Ti1'. The origin of the difference must therefore come from the interactions between Ti1 and Ti1', and Ti3 and Ti3'. For both LDA+U and LDA+SIC, this implies that either the orbital overlaps are too small, or the effective exchange interaction between the orbitals is too small. 
Given the fact that the spin distributions are very nearly identical in DMC and LDA+U in AF1 and AF3 it seems more likely that the LDA+U approximation is too crude and together with the LDA XC functional simply does not have the correct form for the XC potential, even if the orbitals are very accurate. This is also consistent with the large energy gaps obtained with LDA+U. In fact, it is hard to see how a purely local functional, the effective XC functional from the $U$-parameter and the LDA XC functional, can be constructed to simultaneously correct the energy gap and the effective exchange couplings while keeping the atomic moments, spin distributions, and charges unchanged. The same conclusion holds for LDA+SIC, given that the moments and spin distributions in AF1 and AF3 too are very nearly identical to those of DMC.

\section{Conclusions}
We have here analyzed the low-temperature phase the Magn\'eli phase Ti$_4$O$_7$ and its non-trivial magnetic states. We used very accurate Quantum Monte Carlo methods to decisively clarify the nature and energetic ordering of the states.  This clearly demonstrates the applicability of highly accurate QMC methods to magnetic systems with competing low-energy states. We also compared our results from DMC with LDA+U and LDA+SIC in order to gain some insight into the errors generated by LDA+U and LDA+SIC. Our results from DMC shows that LDA+U, LDA+SIC, and also HSE06, give the right states with the right sequence in total energy, while the energy separations are different from those obtained using DMC; especially the FM energy is overestimated by LDA+U and HSE06 relative to the AF states. 
The larger energy difference between the AF3 and AF1 states obtained by DMC compared to the results from LDA+U and LDA+SIC is the effect of correctly including electronic correlations. 
Both LDA+U and LDA+SIC obtain total moments on the Ti$^{3+}$ sites in very good agreement with DMC. However, detailed comparisons between the DMC results, on the one hand, and the LDA+U and LDA+SIC results on the other hand, show that LDA+U and LDA+SIC generate errors in the spin distributions about the Ti$^{3+}$ sites in the ferromagnetic state, while the spin distributions in the AF states are in good agreement with those from DMC. Furthermore, the energy gap for all three states are severely overestimated both by LDA+U and LDA+SIC, in spite of the good agreement in the obtained spin moments between the three methods. This demonstrates that attempts to correct LDA+U and LDA+SIC can in all likelihood not be based on local XC functionals: there appear to be no possibilities within a local XC scheme to simultaneously correct the energy gap and the spin spin distributions while at the same time maintaining the spin moments on the Ti$^{3+}$ sites. This suggests that methods that correct LDA+U or LDA+SIC need to contain more complicated non-local forms of the XC potential.
 
\section*{Acknowledgments}
An award of computer time was provided by the Innovative and Novel Computational Impact on Theory and Experiment (INCITE) program. This research used resources of the Argonne Leadership Computing Facility, which is a DOE Office of Science User Facility supported under Contract DE-AC02-06CH11357. Sandia National Laboratories is a multiprogram laboratory managed and operated by Sandia Corporation, a wholly owned subsidiary of Lockheed Martin Corporation, for the U.S. Department of Energy's National Nuclear Security Administration under Contract No. DE-AC04-94AL85000. AB, LK, JK and PK are supported through Predictive Theory and Modeling for Materials and Chemical Science program by the Basic Energy Science (BES), Department of Energy (DOE). The work by O.H. was supported by the Department of Energy, Office of Science, Division of Materials Science and Engineering. X.Z. was supported by U. S. DOE, Office of Science under Contract No. DE-AC02-06CH11357

\bibliography{Magneli}

\end{document}